# The use of industrial molasses waste in the performant synthesis of few-layer graphene (and its Au/Ag nanoparticles nanocomposites): Photocatalytic and supercapacitance applications

*Kamel Shoueir* [a,b]*, *Anurag Mohanty* [b], *Izabela Janowska* [b]*

[a] Institute of Nanoscience & Nanotechnology, Kafrelsheikh University, 33516, Kafrelsheikh, Egypt

[b] Institut de Chimie et Procédés pour l'Énergie, l'Environnement et la Santé (ICPEES), CNRS UMR 7515-Université de Strasbourg, 25 rue Becquerel 67087 Strasbourg, Franc

*kamel_rezk@nano.kfs.edu.eg, tel. +(33) 0749541481

*janowskai@unistra.fr, tel. +(33) 0689101253

**Abstract**

In view of clean environment, the industry needs to address multiple demands at different levels of production and processes via the sustainable approach including recycling or smart use of produced waste. On the other hand, a development and success of green energy requires the crucial materials synthesized via efficient, sustainable methodology. Herein, we present the green, simple, easily scalable, fast, and highly efficient synthesis of few-layer graphene (FLG) and its composites with Au and Ag nanoparticles using a waste. The FLG synthesis based on the exfoliation of graphite occurs in water in the presence of industrial co-product, molasses, which next shows also performant reductive properties during Ag and Au NPs formation. The decreased size of NPs deposited on FLG indicates the synergetic effect of molasses and FLG, exhibiting the add role of FLG/molasses as metal stabilizer species. The NPs/FLG composites are efficient photocatalysts in degradation of organic contaminant, bisphenol (BisA), in the presence of peroxy-monosulfate (PMS) activator. The Au/FLG (PVDF) and Ag/FLG (PVDF) based electrodes reveal as well relatively high gravimetric capacitance, 205 Fg$^{-1}$ and 729 Fg$^{-1}$. The presented approach is much worthy to be further applied in the synthesis of other layered materials as well as other non-noble supported metallic systems.

1. **Introduction**

Concerning food wastes, some industrial co-products contribute to environmental releases in significant extend. Prevention and recycling or re-utilization are essential strategies [1]. One of such industrial byproducts comes from sugarcane (*Saccharum officinarum L.*) transformation, *i.e.* molasses produced during sugar manufacturing process as thick, dark black liquid, known also as mother syrup [2]. It includes not only different polysaccharides such as glucose (5 % w/v), fructose (5 % w/v), sucrose (45 % w/v), but also heavy metals, nitrogen-containing compounds, and vitamins [3]. It is already used as inexpensive raw material for many industrial chemicals, for example white gum, absolute ethanol, citric acid, lactic acid, and exo-polysaccharide [4].

Graphene has gain a lot of attention owing to its excellent physical and chemical characteristics such as excellent electronic an thermal conductivity, charge transport, high theoretical specific surface area, or mechanical resistance [5]. The often combined properties open many application fields for graphene of few layer graphene (FLG) and their composites [6]. A range of methods has already been developed to produce graphene and graphene based materials, where two main methodologies are developed depending on the envisaged applications, i.e. bottom-up and top-down. The efficient and sustainable synthesis is indeed the factor determining the expected commercial success of these materials. The top-down approach based in general on the exfoliation of graphite provides usually higher yield and relatively lower (but sufficient) quality. The recent trends concern the exfoliation in water either via oxidation of graphite to graphene oxide (GO) via Hummer's method [7] or by using the surfactants [8]. Few apply additionally non standard techniques such the use of kitchen blender [9]. Reduced GO (rGO) obtained via graphite oxidation, exfoliation and finally reduction is the most common graphene-based material but its synthesis has many drawbacks. The synthesis is time-consuming, requires harsh conditions and the reduction step is hardly controllable and difficult. Concerning the exfoliation in the presence of surfactants, several reports proposed the molecular bio-surfactants [10] with some limitation relating to the size of the obtained graphene/FLG flakes and yield/production rate [11]. We have reported previously the easily scalable and efficient exfoliation of layered materials that allowed to get their highly concentrated aqua colloids. The method proposed the use of large, natural bio-surfactant systems (oil/water emulsifiers possesing high hydrophilic-lipophilic balance (HLB)) together with the use of mixing-assisted sonication technique [12]. It resulted in few micro-dimensional graphene flakes when expanded graphite was exfoliated [13]. Although this approach was already highly sustainable, the HLB bio-surfactant system (serum albumin, maltodextrin or agar-agar) needs to be extracted for this purpose. Likewise, the HLB systems undergo modification under sonolysis that producing the large size impurities difficult to be separated (ex. the nano-micro spheres from serum albumin).

Herein, we propose to go a step further with the sustainable synthesis of FLG and its composites with metal NPs by exploitation of important industrial waste, cane molasses, first as bio-surfactant to produce FLG via rapid exfoliation of graphite in water and next as a reducing agent to produce Au and Ag NPs supported over FLG. Beyond the efficient synthesis of the materials, we observe the reduction of NPs size due to the synergetic effect of FLG and molasses during the NPs formation. The Au/FLG and Ag/FLG nanocomposites are evaluated as photocatalyst in oxidative degradation of bisphenol A (BisA) using an activator, and as supercapacitor electrodes. Herein approach fits in a waste to resource strategy for organic contaminants degradation in water, which recently covered several interesting catalysts and advanced processes such as the zing slag (Zn-S) in ozonation of Direct Back 22 or photo-assisted Fenton-like process in the presence of mineral activator [14, 15].

Concerning the preparation of Au/Ag NPs and graphene composites in view of photocatalytic applications, one can find several examples in literature [16]. Some examples deals with the sustainable development principles and use natural species as reducing/stabilizing agent [17]. Quasi all of them contains rGO, which as mention above is obtained via highly disadvantageous processes [18].

**Experimental Section**

1.1. **Chemicals**

Pure graphite powder (<20 μm, synthetic, extra pure), Bisphenol A (BisA), peroxymonosulfate (PMS) activator (known as Oxone, $2KHSO_5 \cdot K_2SO_4$), Gold (III) chloride hydrate (99.9% trace metal basis), Silver nitrate crystal ($AgNO_3$, ACS reagent, ≥99.0%), Poly(vinylidene fluoride) pellets form (PVDF, average Mw 530,000, anhydrous 1-Methyl-2-pyrrolidinone (NMP, ≥99%), L-Histidine (His, ≥99%), Ethylenediaminetetraacetic acid (EDTA, BioUltra, anhydrous, ≥99%), p-Benzoquinone (≥99% (HPLC), Sodium sulfate ($Na_2SO_4$, analytical grade), and Isopropyl alcohol ((Iso), natural, ≥98%) were obtained from Sigma- Aldrich. Sugar cane molasses was obtained from the United Sugar Company of Egypt (USCE). Dialysis membrane tubing with cut-off of 6000 to 8000 Dalton was obtained from Fisher Scientific. All solutions in the present protocol were prepared using double deionized water (DDI) from the Millipore system (18.2 MΩ/cm at 25°C). Any other used reagents were used without any specific pretreatment.

**1.2. Molasses assisted *in-situ* preparation of NPs/FLG**

The accurate weight of concentrated cane molasses (1 g) was diluted with 100 ml of DDI to produce 1 wt. % solution. The specs of the cane molasses batch were as follows: Brix % 81.2, Polarization 40.6%, Water content 22%, and Ash content equal to 4.5%. FLG was obtained by adding 0.4 g of pure graphite into 100 mL molasses solution under sonication probe (Q700 sonicator) and gentle stirring at room temperature for 3 h. The suspension was purified by contact dialysis membrane against one-liter DDI for another 3 h with repeating water change, when necessary, to eliminate most of the residues of polysaccharides. Then, the obtained powder was centrifuged at 10000 rpm with repeated washing using water/ethanol mixture (20/80) followed by drying at 65 °C for 48 h. The Ag/FLG and Au/FLG were obtained by dropwise addition of 1mmol of $AgNO_3$ (or $HAuCl_4.5H_2O$) in 20 mL after completion of FLG formation (FLG/molasses) under stirring at 70 °C for 30 min. The nanocomposites were dried at 65°C for 48 h.

**1.3. Characterization tools**

The particle shape and size of the prepared nanoparticles and nanocomposites were evaluated by HR-TEM, JEOL, 2100, Japan, at the accelerated voltage at 200 kV. The crystal structures were confirmed by XRD with X-ray diffractometer (analytical-X part pro, Cu $k_{α1}$ radiation, λ=1.5404 Å, 45 kV, 40 mA, Netherlands) in the range of $10º ≤ 2θ ≤ 80º$ with step size of 0.02º, and an irradiation time of 0.5 s per step. UV-Vis Spectrophotometer Double beam (Uv-1900i) was used to detect the absorbance of BisA. The FLG morphology was confirmed by FE-SEM, FEG250/Quanta with the acceleration voltage of 30 kV. The scanning was associated with a pendant energy dispersive X-ray (EDX) detector. The Raman spectroscopy was performed using laser excitation at 514.5 nm from the argon-ion laser source (Lab-RAM HR Raman spectrometer). TGA was used to investigate the thermal stability on Perkin Elmer TGA 6, in the range from 0 to 800 °C using air atmosphere with a heating rate of 10 °C/min. XPS spectral investigation was performed on a THERMO MULTI LAB 2000 spectrometer equipped with AlK anode at hɣ= 1486.6 eV. THE CASA XPS program fitted with Gaussian-Lorentzian mix function was used for deconvolution of the spectra.

### 1.4. Photocatalytic and electrochemical investigations

#### 1.4.1. Photocatalytic oxidation of BisA

To study the photocatalytic properties of FLG, Au/FLG and Ag/FLG nanocmposites under visible-light irradiation, an approximately 30 µM stock solution of BisA (pH 7.2) was prepared. To retain the temperature constant at 25 ± 2 °C, a cooling bath was used. A small amount of nanocomposite (10 mg) was then placed in a 50 mL cylindrical Pyrex container with 50 mL of BisA (pollutant model), and a specific amount of BisA (13.14µM) was mixed with PMS (1mM). The admixture solutions were subjected to adsorption-desorption equilibrium via magnetic stirring for 60 min in the dark state before switching on the light and adding an activator (PMS). The admixture was continuously stirred during the experiment, and the pH was altered by using diluted solutions of NaOH (0.1 M) and $H_2SO_4$ (0.1 M) when necessary. A sample volume (approximatively 1 mL) was withdrawn at regular intervals and blended with EDTA, ISO, His, and BQ to slake the remaining free radicals, followed by centrifugation to determine the BisA concentration. The BisA concentration was estimated using Uv-vis double beam at maximum absorbance ($\lambda$ = 276 nm).

#### 1.4.2. Electrochemical evaluation test

The electrochemical properties of the prepared Au/FLG and Ag/FLG were investigated in 1 M $Na_2SO_4$ aqueous solution over a voltage window of 0 - 1V in a three-electrode system and measured with a Biologic SP-150 Potentiostat/galvanostat. The three-electrode system consisted of Pt coil counter electrode, calomel reference electrode (Ag/AgCl) and working electrode (Au/FLG or Ag/FLG). The Au/FLG and Ag/FLG electrodes were conducted as electrodes for supercapacitor using galvanostatic charge-discharge (GCD), cyclic voltammetry (CV), and electrochemical impedance spectroscopy (EIS) techniques. GCD experiments were measured at different specific currents (current density: 0.46 – 4.6 A g$^{-1}$). The CV measurements were performed at various scan rates (10 – 100 mV s$^{-1}$). The EIS parameters were derived using the EC-Lab V11.33 software. The EIS measurements were evaluated within the frequency range from 10$^{-2}$ to 10$^{5}$ Hz with a sinusoidal perturbation amplitude of 10 mV.

**1.4.2.1. Preparation of working electrodes**

The manufacturing of working electrodes were as follows: 20 mg of Au/FLG or 20 mg of Ag/FLG, and 2 mg polyvinylidene fluoride (PVDF) were thoroughly mixed. The produced homogeneous slurry was obtained by adding the admixtures to N-methyl-2-pyrrolidone (NMP, 0.3 mL). Then, the resultant suspension was coated onto a 0.1 mm thick stainless-steel sheet (1 cm$^2$) with mass loading of 2.4 mg cm$^{-2}$. Finally, the designed electrodes were dried in the oven at 50 °C for 12 h.

**2. Results and discussion**

**2.1. Synthesis of FLG and Au/FLG, Ag/FLG nanocomposites**

As described in the Introduction a main goal of the current work was the smart use of molasses co-product waste in the preparation of FLG and its composites with Au and Ag NPs via facile, cost-effective, green and easily scalable method. Molasses is a mixture of different molecules which together have a high hydrophilic-lyophilic balance (HLB) character. Fig. 1S depicts most of these molecules including different polyphenols (black color), amino acids (blue color), and polysaccharides (red color) according to GS4 Methods of International Commission for Uniform Methods of Sugar Analysis (ICUMSA) [19]. The importance of high HLB character was highlighted in our previous work [12]. During the exfoliation, the lyophilic moieties interact with the hydrophobic surface of graphite, while the hydrophilic units interact with water. These interactions help to overcome the van der Waals forces between graphite sheets and next to stabilize the obtained FLG in water. As a product the FLG flakes with some adsorbed molasses molecules remained after dialysis separation step are obtained.

Otherwise, the molasses is a powerful reducing agent; it required a few minutes to reach a complete reduction of Ag and Au precursor salts ($AgNO_3$ and $HAuCl_4.5H_2O$) without using an extra reducing agent. The potent groups in molasses are highly viable to reduce metal salts to metallic nanoparticles as it is a case of some natural extracts ( black currant extract: [20], leaf extract [21]).

We reveal for the first time that the formation of NPs in the presence of the FLG leads to the FLG supported metal which exhibits higher dispersion with a lower average size of NPs compared to the NPs obtained only with molasses. It indicates that molasses over the FLG surface plays an add role of metal stabilizer.

The as-prepared FLG and its nanocomposites are characterized in term of morphology (particle shape, size), crystallinity, chemical structures, and thermal stability.

### 2.2. Structural/morphology features

Fig. 1 illustrates the TEM micrographs of FLG flakes and of the nanocomposites (Au/FLG and Ag/FLG). According to the statistical analysis, the obtained FLG are few μm sizes flakes, Fig. 1a, with 7 sheets in average. The number of sheets can be found by counting the slightly curved edges within a given flake, and in Fig. 1 (b, c) one can observe the flakes containing 4 and 9 sheets. The micrographs of the nanocomposites confirm the existence of crumpled flakes with dimensions greater than 500 nm as well and the presence of supported metal NPs. The nanoparticles of AuNPs (Fig. 1 d-f) and AgNPs (Fig. g-i) are relatively well distributed over the FLG surface. They demonstrate a spherical shape with a size of few nm, slightly larger for AuNPs. The average size of Au NPs is around 8-12 nm with the highest population at c.a. 10 nm, vs. 5-12 nm for AgNPs with the highest population at c.a. 7 nm. The corresponding histograms created after randomly selected hundreds of particles and analyzed by ImageJ program are presented in Fig. 1(j, k). Several NPs of Au have the size bigger than 12 nm (up to 18 nm), whereas in the case of Ag the NP size bigger than 12 nm is rare and accidental going in tandem with the fact that the loading of the Au is c.a. twice higher compared to the Ag (7% vs./ 3%). The relatively small size of NPs is related to the reducing and stabilization action of molasses used during their preparation in synergy with FLG. The presence of molasses and FLG during the preparation of NPs reduced their size per half. The average size of Ag and Au NPs obtained only with molasses was detected to be as high as 16-21 and 14-24 nm with a maximum size reaching around 30 nm (Fig. S2, S3).

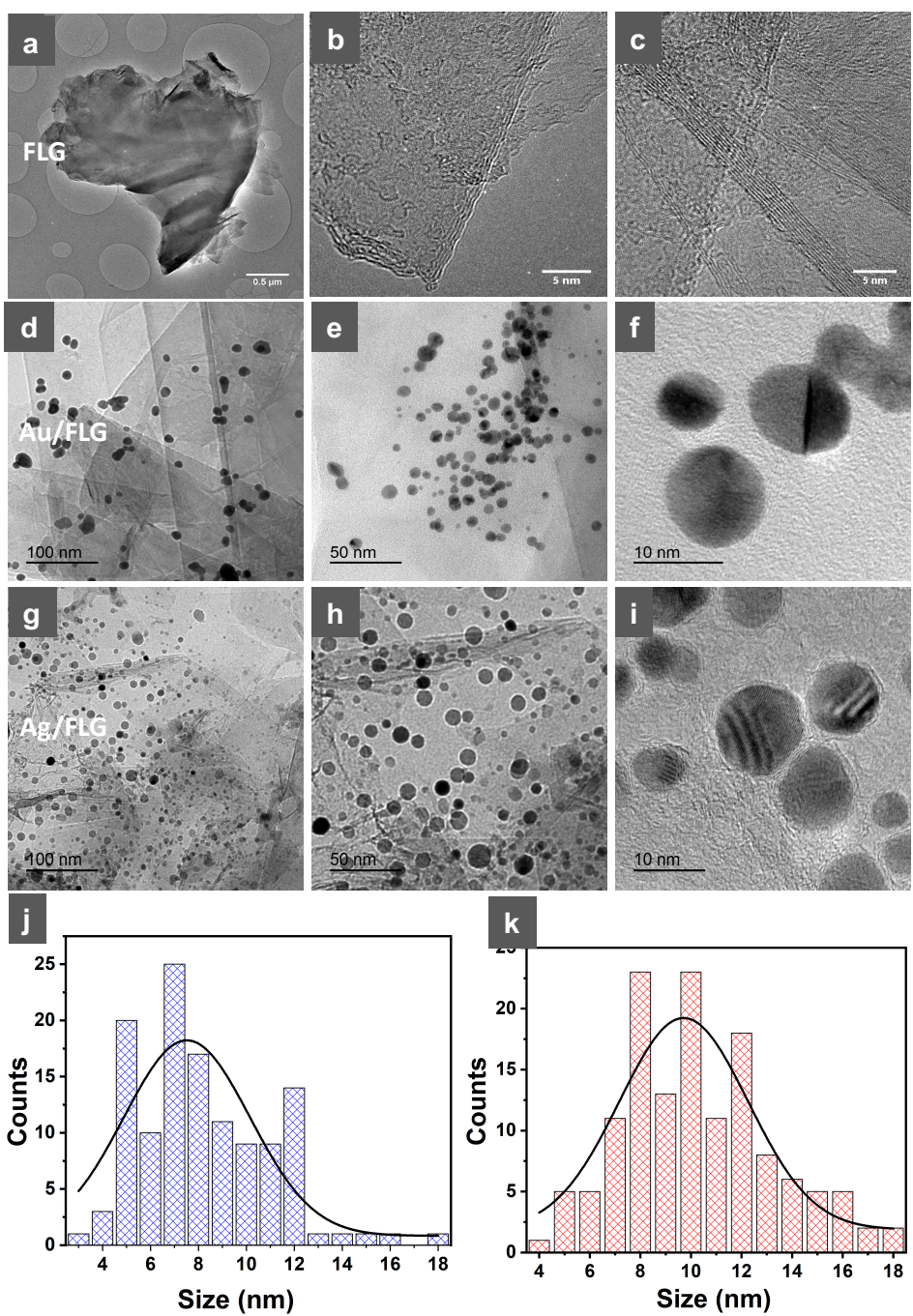

**Figure 1.** TEM micrographs of (a) aggregate of FLG flakes, (b) FLG flake containing 4 sheets (c) 9-sheets FLG flakes, (d, e) Au NPs over FLG, (f) magnified HR-TEM of Au/FLG, (g, h) Ag NPs over FLG, (i) HR-TEM of Ag/FLG revealing high crystallinity of NPs, (j, k) histogram of Ag/FLG and Au/FLG size distribution based on approximately 100 particles.

It has to be underlined that the role of molasses as stabilizing agent for metal phase is clear. Contrary to the GO (rGO), which is highly substituted by oxygen groups on the surface, FLG has no or negligible amount of such functionalities on the surface. Consequently, the preferential sites for metal attachment and stabilization are the FLG edges rich in dangling bonds [22]. Here one can see that such preferential on-edges deposition does not exist.

To have a more general view of the NPs/FLG morphology, FE-SEM analysis was also conducted. Fig. 2 shows aggregates of the synthesized FLG flakes and FLG flakes decorated by AuNPs and AgNPs. The morphology of both nanocomposites is slightly different. It seems that FLG decorated with AgNPs aggregates exhibit flatter and more homogenous features. It forms more compact aggregates during drying of the sample prior to the SEM analysis compared to the Au decorated FLG flakes. Likewise, the presence of bulky Ag/particles aggregates can be detected (Fig. 2 c, d). In Au/FLG (Fig. 2 e,f) the bigger average size of NPs, as also confirmed by TEM, makes FLG aggregate less compact due to the weaker inter-layer interactions. In this case, Au NPs make an efficient spacer for the FLG flakes.

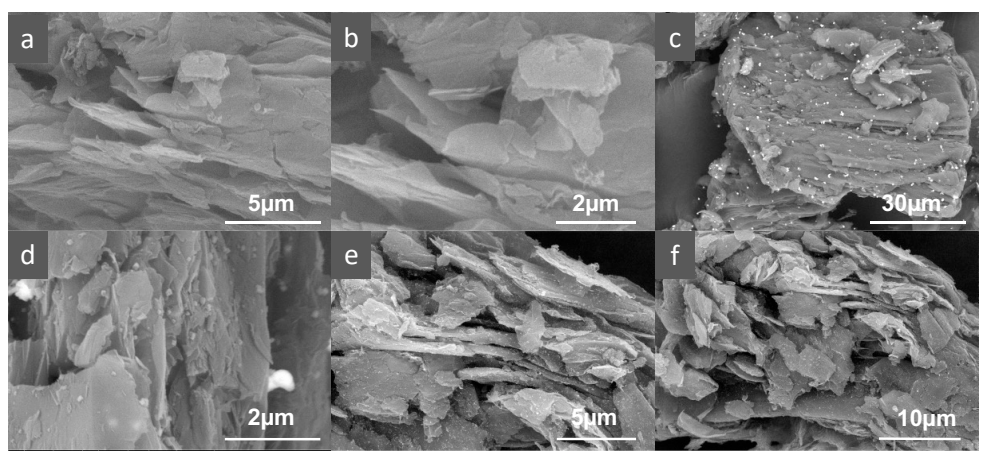

**Figure 2.** FE-SEM micrographs of (a,b) synthesized FLG, (c,d) Ag/FLG, and (e,f) Au/FLG nanocomposites.

## 2.3. Crystallinity via thermal stability, XRD, and Raman spectroscopy

TGA was performed to outline the thermal stability of FLG and FLG decorated with AgNPs and AuNPs, and the obtained data are illustrated in Fig. 3a. In the case of FLG and somehow in Au/FLG, a slight degradation occurs already between 200 and 500°C and can be attributed to the decomposition of molasses residue. Above this temperature, a degradation of FLG begins. As expected a decomposition of FLG in the composites starts at a lower temperature than the decomposition of FLG alone. The maxima of oxidation peaks determined from TGA derivatives are 641, 623, and 592 °C for FLG, Au/FLG, and Ag/FLG. This phenomenon confirms the presence of NPs acting as an oxidizing catalyst during the combustion process. The relatively low combustion temperature in the case of Ag/FLG despite low size and amount of supported NPs confirms the presence of Ag aggregates and its compact structure. The compact structure and high thermal conductivity of well-crystalized FLG induces the formation of hot spots. Similar phenomena can be observed in the polymer composites containing such kinds of FLG flakes at higher loading [23]. Overall, pure FLG and FLG nanocomposites led to the excellent resistance towards thermal oxidation up to 600 °C due to the high crystallinity with only slight mass loss. The TGA analysis confirmed also the % weight of Au and Ag in the composites.

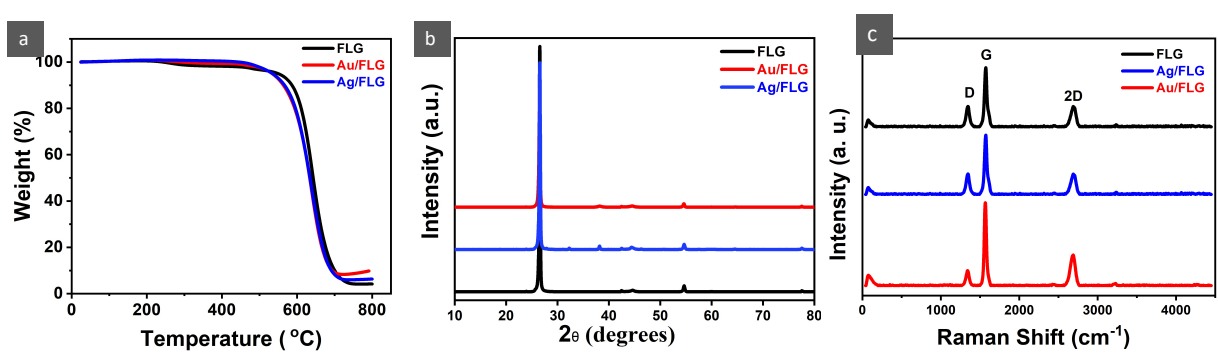

**Figure 3.** (a) TGA, (b) XRD, and (c) Raman spectra of FLG, Au/FLG, and Ag/FLG.

The XRD and Raman spectroscopy analysis revealed the same identity of FLG in the three samples. A clear and relatively thin diffraction peak of a hexagonal graphitic structure is observed at 2θ=27º in XRD pattern as a signature of well-crystallized FLG, Fig. 3b.

In the Raman spectra, Fig. 3c. the D-band is attributed to the disorder-induced scattering, which is mainly caused by the defects of disordered graphite hexagonal symmetry. On the contrary, the strong G band at around 1580 cm$^{-1}$ is assigned to the vibration of sp$^2$-hybridized carbon atoms. The statistically calculated relatively low $I_D/I_G$ from 0.18 to 0.74 for FLG, Au/FLG and Ag/FLG, confirms the highly ordered crystalline structure of FLG [24].

## 2.4. Chemical composition

The XPS spectra of FLG, Au/FLG, and Ag/FLG are presented in Fig. 4. The XPS C1s core-level spectrum reveals the characteristic signals of crystalline carbon in FLG, Au/FLG, and Ag/FLG. The C1s peak deconvolution results in six main components corresponding to C=C, C-C(R), C-OR, C-O, C=O, and O-C=O groups as defined in Table 1 [25].

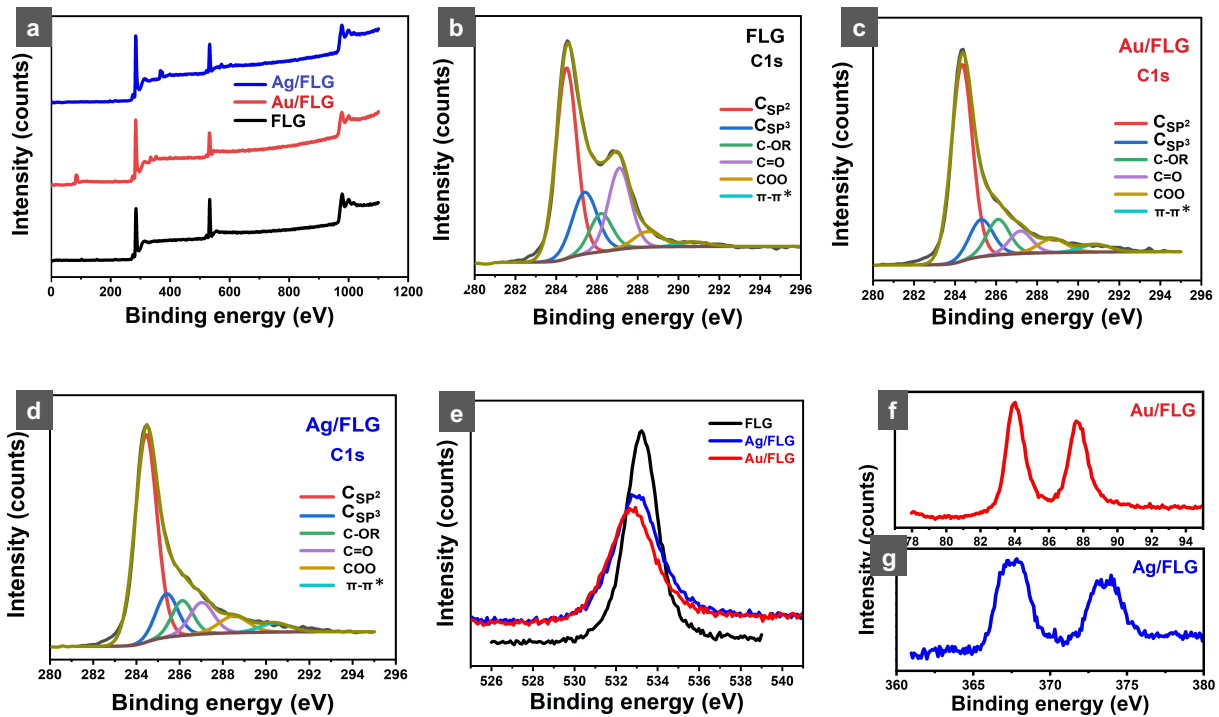

**Figure 4.** XPS spectra of FLG and composites (a) general spectra, (b) C1s spectra of FLG, (c) C1s spectra of Au/FLG (d) C1s spectra of Ag/FLG (e) O1s spectra, and (f) Au/FLG 4f, g) Ag/FLG 3d.

**Table 1:** XPS data of FLG and its Au and Ag composites.

| Samples | C $sp^2$ (284.4 ± 0.2 eV) | C $sp^3$ (285.4 ± 0.2 eV) | C-OR (286.2 ± 0.2 eV) | C=O (287.1 ± 0.2 eV) | COO (288.5 ± 0.2 eV) | π - π* (290.7 ± 0.3 eV) | C/O ratio |
|---|---|---|---|---|---|---|---|
| FLG | 45.2 | 17.7 | 9.8 | 20.9 | 4.9 | 1.5 | 4.2 |
| Au/FLG | 58.7 | 12.3 | 9.4 | 9.7 | 6.4 | 3.5 | 6 |
| Ag/FLG | 59.4 | 12.7 | 11.6 | 7.4 | 5.6 | 3.2 | 5.9 |

According to the XPS spectra, the FLG has a higher non-graphitic contribution than its composites have. First, the oxygen content as determined by the area of O1s and C1s peaks from general spectra is greater. This excess of oxygen is reflected by the strong signature of C=O type groups, which seem to be much more abundant in FLG compared to the nanocomposites. Likewise, the amount of $sp^3$ C is enhanced in the FLG sample being in accordance with TGA analysis. These results clearly indicate the presence of the molasses residue over FLG surface. On the other hand, the absence of molasses residue (or negligible amount) in the composites confirms that the molasses groups over FLG were utilized for attachment and reduction of metal precursors.

Besides XPS spectroscopy, the energy dispersive X-ray (EDX) was conducted to confirm the chemical composition. Table 2 displays the elemental analysis and the atomic weight (%) considering C, O, and metal. The presence of C correlates to the main element of both nanocomposites, which is FLG. Also, the AuNPs are found to contribute at 7.6 wt. % compared to 84.9 wt. % of C and 7.6% of O, while AgNPs were found at 2.2 wt. % compared with 88.8 wt. % C and relative oxygen level up to 9.0%.

**Table 2:** Elemental analysis based on EDX data for Au/FLG and Ag/FLG.

| Composition | Element | Atomic weight (%) |
|---|---|---|
| Au/FLG | C K | 84.9 |
|  | O K | 7.6 |
|  | Au L | 7.6 |
| Ag/FLG | C K | 88.8 |
|  | O K | 9.0 |
|  | Ag L | 2.2 |

The much lower % of Ag (c.a. 2%) compared to Au (c.a. 7%) is reflected as well in TGA analysis. It is worthy to recall that the smaller content of Ag NPs fits with higher dispersion/lower size of NPs compared to Au NPs.

**2.5. Photoluminescence**

Fig. 5 illustrates the normalized Photoluminescence (PL) spectra of FLG, Au/FLG, and Ag/FLG. The excitation peak centers for pure FLG and Au/FLG appear at 276.3 nm, while for Ag/FLG it has been marginally shifted to 276.5 nm. It can be observed that emission peak centers migrated towards higher wavelengths in Au/FLG and Ag/FLG compared to FLG, i.e., to 335.7 and 336.5 nm, vs. 322.5 nm.

Even after the close excitation energy values, the addition of NPs drastically affected the emission wavelengths especially in the Au/FLG sample. A substantial decrease in PL emission intensities from pure FLG to Au/FLG and Ag/FLG implies a direct free path of charge carriers, followed by a low recombination rate. The photoelectrons generated by the Au or Ag NPs are transported to the surrounding atmosphere with the aid of FLG nanosheets indicating the efficient separation of photogenerated electron-hole pairs. One can clearly see that since in the case of FLG, the emission PL intensity is much higher compared to the excitation peak, in the composites, the emission is strongly weakened (especially in Au/FLG). Such PL quenching conclusions in photoactive materials such as functionalized CN were strengthened via photocurrent measurements [26].

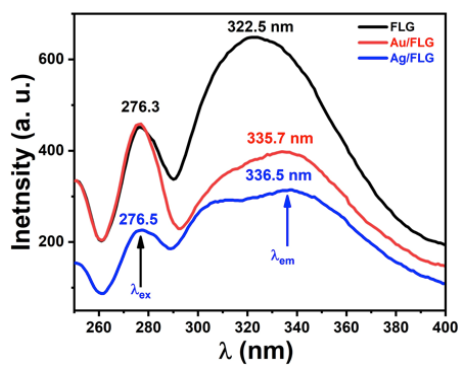

**Figure 5.** Photoluminescence (PL) of pure FLG and that decorated with AuNPs (Au/FLG) and AgNPs (Ag/FLG).

### 2.6. Photocatalytic efficiency in bisphenol-A (BisA) oxidation

Fig. 6 represents the absorption and photocatalytic efficiency of the prepared FLG and nanoparticles supported on FLG to evaluate activity in the oxidation of BisA at pH=7 with or without an activator. BisA was used as an organic model. The tests effectuated in the dark, where the only adsorption of BisA can take place, showed that FLG-based nanocomposites could not remove or degrade the organic compound even after long time (Fig. 6a). The removal efficiency of BisA in the dark or under light without PMS was negligible. The photocatalytic efficiency of FLG and the nanocomposites with the aid of visible-light irradiation was sharply improved in the presence of PMS as a stimulator. They show a pronounced decrease in the BisA concentration as shown in UV-Vis spectra (Fig. 6b-d). After being subjected to visible light, precisely 100, 90, and 65 min are sufficient to achieve good catalytic activity of 57.23, 79.74, and 97.75% for FLG, Au/FLG, and Ag/FLG.

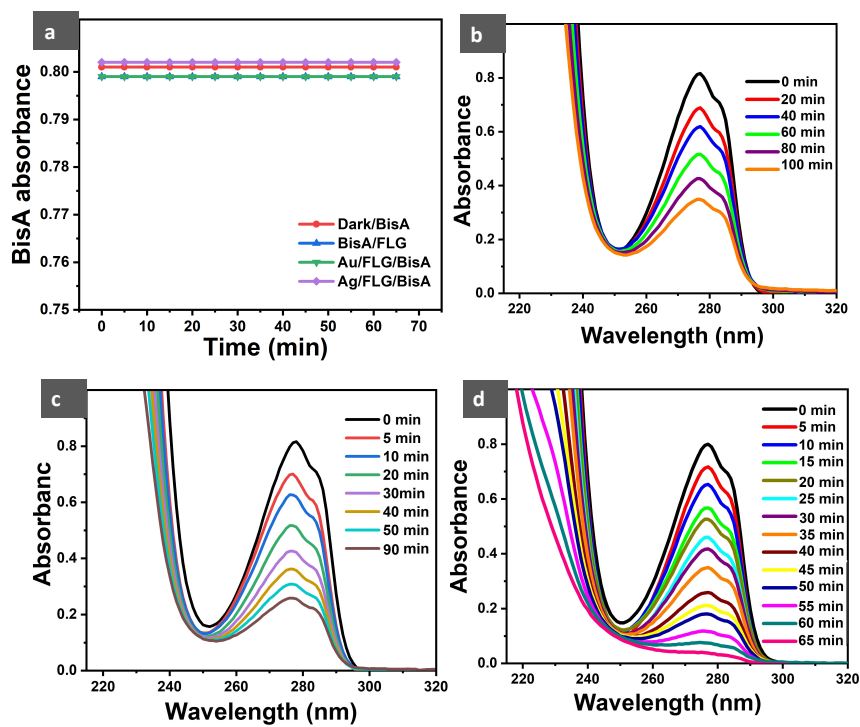

**Figure 6.** (a) absorbance of BisA using different systems under dark conditions with absence of PMS, UV-vis spectrum of (b) FLG, (c) Au/FLG, and (d) Ag/FLG [$C_0$ = 21.9 μM, PMS = 1.0 mM, Dosage = 10mg, pH = 7.2, Temperature = 25 $^0$C, Volume = 50ml]. Error bars depicted the relative standard error after the mean of three replicates.

Concerning the two nanocomposites, Ag/FLG is more efficient in BisA removal than Au/FLG, while the latter is incomparably more performant than FLG. The activity of the nanocomposites is attributed to the formation of radicals which catalyze the degradation of organic dye (BisA). The relatively higher decomposition rate within a limited time, especially for Ag/FLG is similar or higher than many other reported photocatalysts (Table 3). It is difficult however to make a direct comparison between different systems due to the variable conditions used during the tests. We then limit the choice of examples in table 3 to the few best materials with possibly closest conditions to ours and the ones including graphene (rGO). Worthy to note is the extremely low concentration of the catalyst we used (0.01 g/L).

**Table 3**: Comparison between Ag/FLG nanocatalyst and some other photocatalytic systems for BisA degradation in the presence of PMS activator.

| Catalyst | Conc of BisA (μm) | General conditions | Efficiency | Time (min) | Ref |
|---|---|---|---|---|---|
| FeCoS-4@N-rGO | 1.5 | Catalyst dosage = 0.03g/L, PMS = 0.87mM, pH = 7.0 | 98.3% | 20 | [27] |
| N-doped rGO membrane | 408.2 | Catalyst dosage = 0.03g/L, PMS = 1mM, pH = 7.0 | 80% | 120 | [28] |
| Co/CoOx@NC/ rGO | 87.6 | Catalyst dosage = 0.1g/L, PMS = 2.19m, M pH = 6.8 | 100 | 60 | [29] |
| Fe Al-LDH | 20 | Catalyst dosage = 0.2g/L PMS = 0.2 g/L, pH = 5.8 | 93% | 60 | [30] |
| Mn-Fe LDO | 20 | Catalyst dosage = 0.4g/L PMS = 1.5 mM, pH = 7.0 | 100% | 50 | [31] |
| MIL-88A(Fe)/MoS$_2$ | 20 | Catalyst dosage = 0.1g/L PMS = 0.5 mM, pH = 7.0 | 98.2% | 60 | [32] |
| NiZn@N-G-900 | 20 | Catalyst dosage = 0.2g/L PMS = 0.244 mM, pH = 6.5 | 100% | 80 | [33] |
| Ag/FLG | 21.9 | Catalyst dosage = 0.01g/L PMS = 1.0 mM, pH = 7.2 | 97.75% | 65 | This study |

BisA degradation curves are plotted using a pseudo-first-order kinetic model represented by $\ln\left(\frac{A}{A_0}\right) = -k_{app}t$, where $k_{app}$ is the pseudo-first-order model constant (min$^{-1}$) calculated from the slope of the curves drawn as ln (A/A$_o$) as a function of irradiated time, Fig. 7a. The calculated constants ($k_{app}$) are 0.0063, 0.016, and 0.038 min$^{-1}$ for FLG, Au/FLG, and Ag/FLG. The latter vastly promotes the photon-induced degradation, and together with the PL investigations, suggests better electrons migration rate and separation of charge carriers (Fig. 7b).

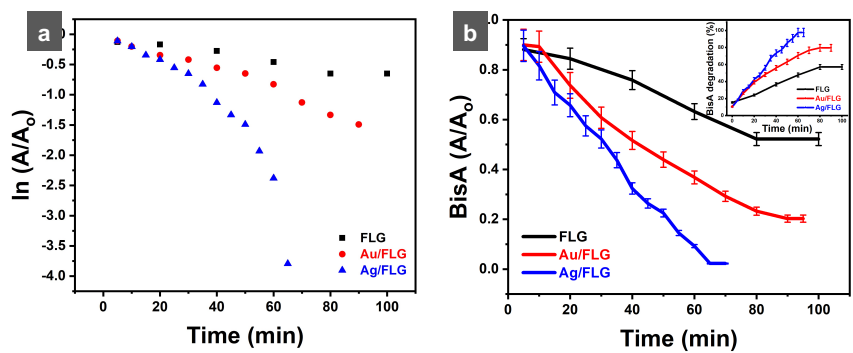

**Figure 7.** (a) BisA degradation kinetics as a function of time and (b) Photodegradation of BisA (insert: oxidation percentage (%)) for FLG based nanocatalysts.

Concerning metal-free FLG, it was reported elsewhere that different carbon defects as exposed edges, heteroatoms, vacancy sites, and few sheets in FLG are preferable for the catalytic activation of PMS [34]. In our case, the presence of molasses residue on the surface of FLG can also somehow modify PMS activation and hinder the Bis A adsorption. In the composites, the molasses is consumed for the stabilization and reduction of NPs but still the adsorption of BisA is contestable. Concerning Au and Ag NPs, it is highly probable that the Surface Plasmonic Resonance (SPR) occurs [35], increasing the charge carrier's rate owing to the strong local electromagnetic fields [36]. The SPR would facilitate the transfer of electrons from the surface plasmon (SP) present at a higher level to the lower level conduction band (CB) of FLG [37].

The difference in photocatalytic degradation activity between both nanocomposites in this study can be related to the variable dispersion of both metals over FLG, including their amount and NPs size (which consequently would effect SPR effect).

### 2.6.1. Quenching mechanism of PMS activation

The mechanism of BisA removal is considered mostly as radical oxidation. As illustrated in Fig. 8, and according to the investigations below, the formed $SO_4^{\bullet-}$ and $\bullet OH$ are prominent active radicals in the catalyst-mediated PMS activation systems [38].

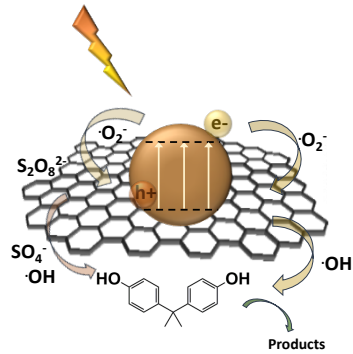

**Figure 8.** Photooxidation of BisA mechanism over FLG and its Au and Ag-based nanocomposites.

The main chemical reaction can occur as follows [39]:

$$SO_4^- + H_2O \rightarrow OH\cdot + SO_4^{2-} + H\cdot \quad (1)$$

$$SO_4^- + OH^- \rightarrow OH\cdot + SO_4^{2-} \quad (2)$$

The powerful superoxide radicals $O_2^{\cdot-}$, $\cdot OH$, and non-radical $^1O_2$ (ROS) species can also be produced as illustrated in equations (3-5) mainly during PMS activation. The presence of different radicals and ROS are investigated here by implementation of some typical scavengers, such EDTA, isopropyl alcohol, histidine (His), and 1,4-benzoquinone (BQ).

$$HSO_5^- + H_2O \rightarrow HSO_4^- + H_2O_2 \quad (3)$$

$$OH\cdot + H_2O_2 \rightarrow H_2O + HO_2^\cdot \quad (4)$$

$$HO_2^\cdot \rightarrow H^+ + O_2^- \quad (5)$$

EDTA was used to quench $\cdot OH$ and $SO_4^{\cdot-}$ and isopropyl was utilized as a radical scavenger for $\cdot OH$ while having also lower rate constant for $SO_4^{\cdot-}$[40]. Concerning His and BQ, their high analytical performance towards $O_2^{\cdot-}$ radical [41] and singlet oxygen $^1O_2$ was assessed [42]. As shown in Fig. 9, the oxidation degree of BisA in the presence of scavengers also varies slightly depending on the sample, while the general tendency remains the same. One can see, that in all samples, a degradation of BisA is inhibited in a significant manner when EDTA and isopropanol are added. This confirms the important role of $\cdot OH$ and $SO_4^{\cdot-}$ ( EDTA being stronger inhibitant compared to isopropanol due to its sensitivity towards $\cdot OH$ and $SO_4^{\cdot-}$ and higher rate constant ($k_{SO_4^{•-}}$ = 2.5 × 10$^7$ M$^{-1}$ s$^{-1}$, $k_{•OH}$ = 9.7 × 10$^8$ M$^{-1}$ s$^{-1}$) [43]). On the other side, the presence of 3mM of BQ almost does not influence the BisA oxidation, and only the scavenging effect of His is clearly visible. This latter indicates that ROS species take some part in the BisA oxidation. Similar results were reported for instance for g-C$_3$N$_4$ materials [44]. In Co/V-g-C$_3$N$_4$ system the formation of •OH and O$_2^{•-}$ were confirmed via ESR technique [45].

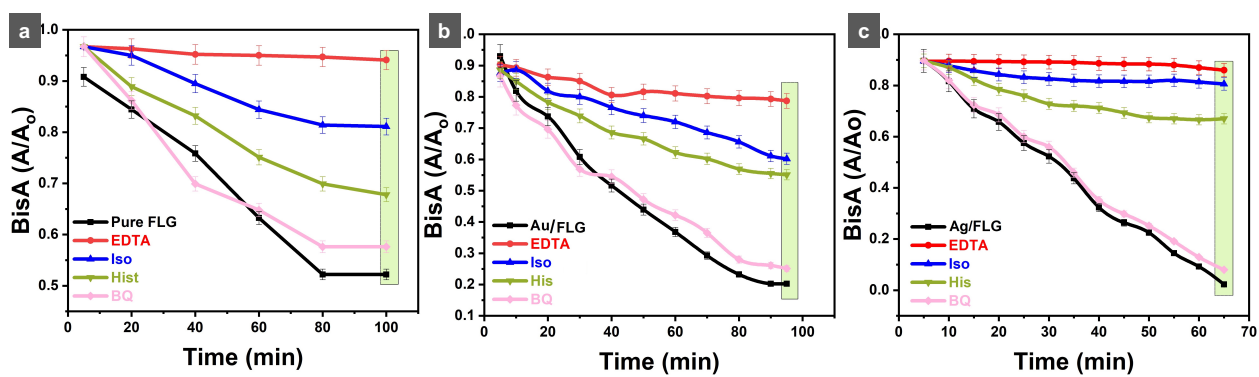

**Figure 9.** Radical scavengers effect on BisA degradation as a function of time using FLG and FLG based nanocomposites: (a) FLG, (b) Au/FLG, and (c) Ag/FLG.

### 2.6.2. Stability and longevity performance

In literature, metallic nanoparticles with excellent light absorption capacity have been reported extensively [46]. For any practical application, the stability of the catalytic system is critical taking into account the potential photocorrosion driven loss of activity [47]. With this in mind the reusability of the samples have been explored up to six runs.

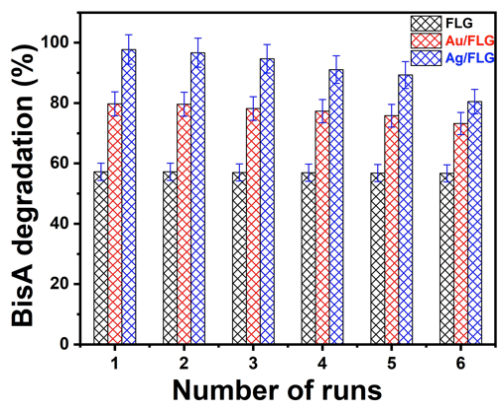

**Figure 10.** Reusability by cyclic experiments up to six runs of FLG, Au/FLG, and Ag/FLG nanocomposites.

After each experiment the samples were thoroughly washed with pure water and then vacuum-dried. Fig. 10 demonstrates that the photocatalytic behavior of pure FLG remains constant even after six runs in the presence of PMS, without any changes. On the other hand, the degradation activity of Au/FLG and Ag/FLG decreases systematically by 6.54 % and 17.25 % after 6 runs.

The stability of photocatalytic activity of Ag/FLG is lower, which can be related to different dispersion of both metals and the nature of the metal itself. SEM was used to depict the stability of FLG-based metallic nanoparticles morphology (Fig S4, S5).

## 2.7. Electrochemical performance of Au/FLG and Ag/FLG nanocomposites

The electrochemical performance of Au/FLG and Ag/FLG was studied by cyclic voltammetry (CV), galvanostatic charge/discharge (GCD), and complemented by electrochemical impedance spectroscopy (EIS) techniques. The measurements including Au/FLG (PVDF) and Ag/FLG (PVDF) as active materials were performed in aqueous electrolyte $Na_2SO_4$ (1 M) using a three-electrode system. The specific capacitance of the electrodes ($C_{sp}$) was measured using the following equation (Eq. 1).

$$C_{CV} = \frac{Q}{ms\Delta V} \qquad \text{Eq. 1,}$$

Where $C_{CV}$ is the specific capacitance, Q is stored charge in coulombs (equals half the integrated area of the corresponding CV curve), m is the weight of the electrode (g), s is the scan rate, ΔV is the potential window. Fig. 11 (a, b) demonstrates the CV curves of Au/FLG and Ag/FLG electrodes in the potential range from 0 to 1 V at different scan rates: 10-100 mV/s. The CV curves of Au/FLG and Ag/FLG electrodes at different scan rates have the shape quite far from the rectangular -a signature of an electrical double layer (EDL) supercapacitance. It indicates the significant contribution of pseudocapacitance (Ps) related to faradic, redox reactions. The highest EDL characteristics are only demonstrated by Ag/FLG electrode at the lowest scan rate.

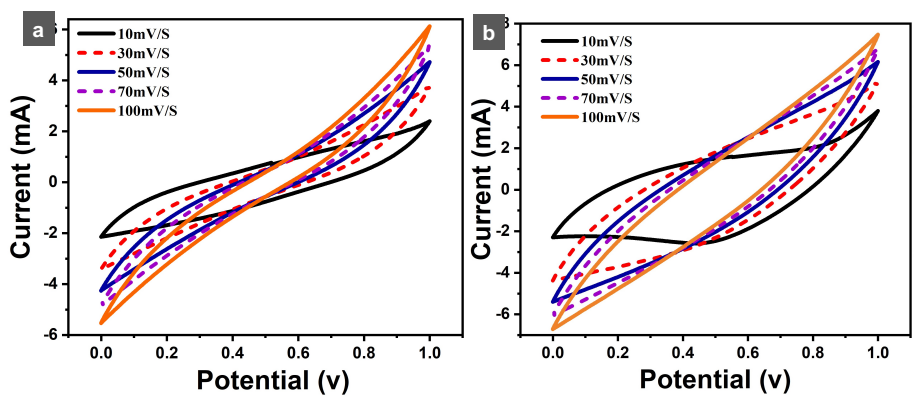

**Figure 11.** CV curves of (a) Au/FLG, (b) Ag/FLG at different scan rates.

To determine the performance of any material for use as a supercapacitor, galvanostatic charge/discharge measurements are essential to be elucidated. The galvanostatic charge/discharge curves at different currents from 1 to 10 mA for Ag/FLG and Au/FLG electrodes are quasi-triangular, Fig. 12 (a, b), indicating relatively fast transfer of charge and electrical conductivity while clear IR drop is visible [48]. The charging-discharging process is slightly faster in Au/FLG composites.

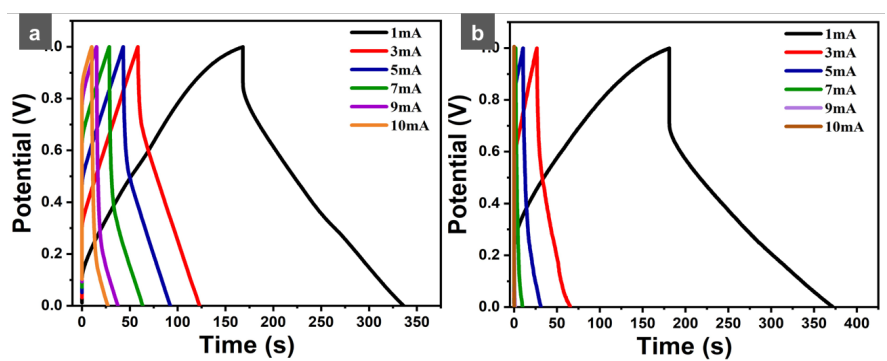

**Figure 12.** GCD curves of (a) Ag/FLG and (b) Au/FLG at currents from 1 – 10 mA.

Fig. 13 (a) elucidates the relation between the specific capacitance and the currents of the prepared electrodes. The electrode materials showed a similar decrease tendency of capacitance with the discharge current rates increase. The lowering of specific capacitances with rising currents is possibly accredited to the insufficient ionic contact to complete the faradic reaction. This can be explained by reducing the diffusion of electrolyte ions with a time constraint at high scan rates. On behalf of charge storage, only the external active surface can be operated leading to a decrease in the electrochemical process of electroactive materials. The specific capacitance was calculated at different currents using the below equation (Eq.2).

$$C_{sp} = \frac{I}{m} \frac{\Delta t}{\Delta V} \qquad \text{Eq. 2,}$$

where I is the discharged current (A), m is the mass of active material (g), Δt is the discharged time (s), and ΔV is the potential window (V). The specific capacitance of 205 F/g was achieved for Au/FLG nanocomposite while 729 F/g for Ag/FLG at the current density of 0.46 A/g. This difference can be attributed to the higher conductivity of the Ag/FLG electrode, as seen from EIS measurements.

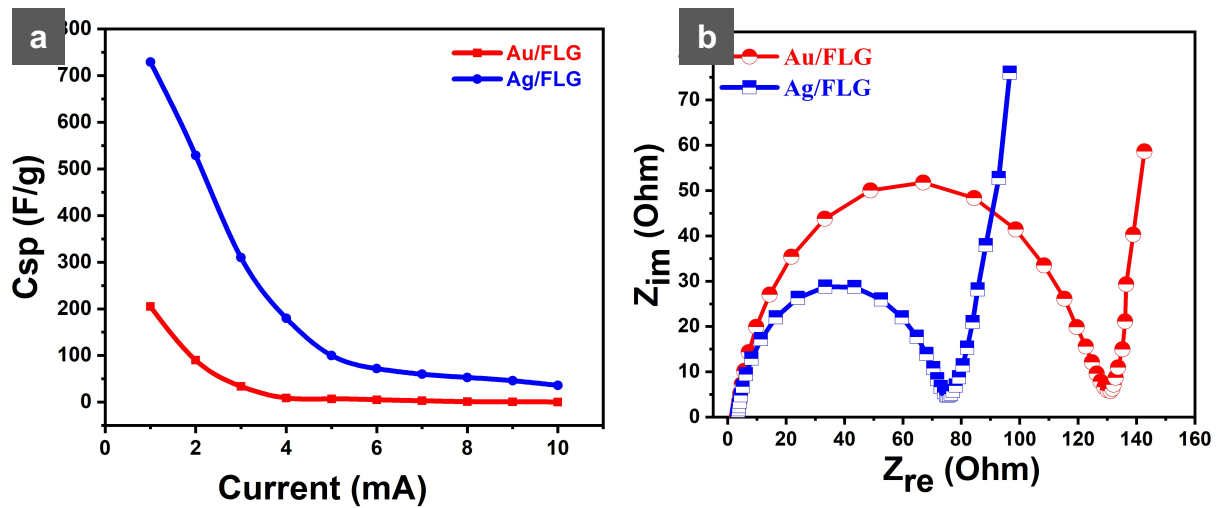

**Figure 13** (a) $C_{sp}$ according to GCD of Au/FLG and Ag/FLG electrodes at different current densities, and (b) Nyquist plots of Au/FLG and Ag/FLG electrodes.

Further, the Au/FLG nanocomposite showed an energy density of 28.5 Wh/Kg at a power density of 351.4 W/Kg, while Ag/FLG nanocomposite showed an ultra-high energy density of 101.25 Wh/Kg at a power density of 243 W/Kg for the current density of 0.46 A/g as shown in Fig. 13 (a). Table 4 includes the present as well as some other data obtained for different systems containing FLG. It has to be outlined that the presented comparison has only approximative character since the data are not related only to intrinsic properties of the materials but also to the form of FLG material. The last point is extremely important since non-3D macronized FLG deposited simply as a film will demonstrate negligible capacitance due to the loss of the accessible electrochemical surface area; the loss caused by inter-flakes π-π stacking interactions. The best example of appropriate FLG macronization is the vertically aligned FLG-Carbon composite. It exhibited high /very high gravimetric/volumetric supercapacitance due to the vertical alignment of FLG, ultramicroporosity and high density despite low surface area [49]. The shaping of the synthesized here nanocomposites are not a goal of this work but it is worthy to note that Ag/FLG exhibits very high capacitance compared to Ag/3D graphene [50] despite much lower Ag content and non-structured graphene. High capacity in our system can be attributed first of all to the well-crystallized and consequently well conductive FLG flakes, as well as the presence of conductive PVDF.

**Table 4.** Comparison of electrochemical performance of different FLG-based materials.

| Active material | Electrolyte | Substrate | $C_{sp}$ (Fg$^{-1}$) at (A/g) | ED (Wh Kg$^{-1}$) | PD (W Kg$^{-1}$) | Ref |
|---|---|---|---|---|---|---|
| Peanut shell-FLG | 1M H$_2$SO$_4$ | Glassy Carbon | 186 (0.5) | 58.125 | 37.5 | [51] |
| FLG-Carbon | 0.5M H$_2$SO$_4$ | Glassy Carbon | 322 (0.5) | 54 | 269 | [49] |
| Nitrogen-FLG | 6M NaOH | Graphite paper | 277 (1.0) | - | - | [52] |
| PANI: PSS/Fe-FLG | 0.1M KCL | Gold (Au) | 659.2 (1.0) | 44.9 | 1750 | [53] |
|  | 0.1M KCL | Gold (Au) | 768.6 (1.0) | 52.3 | 350 |  |
| HCSs FLG-framework | 2M KOH | Glassy Carbon | 561 (0.5) | - | - | [54] |
| FLG-Sheets* | 6M KOH | Pt. Foil | 135 (0.1) | - | - | [55] |
| RuO$_2$–FLGs | 1M H$_2$SO$_4$ | Si wafers | 650 | 57.5 | 23000 | [56] |
| MnO/FLG | 1M Li$_2$SO$_4$ | Ni substrate | 778.5 (0.5) | 81 | 292 | [57] |
| S, N-FLG | 6M NaOH | Graphite sheet | 298 (1.0) | 15 | 300 | [58] |
| Ag/3D graphene (10% of Ag) | 1M KOH | Graphite | 554 (1.0) | - | - | [50] |
| Au/FLG (PVDF) | 1M Na$_2$SO$_4$ | Stainless-steel | 205 (0.5) | 20.3 | 356.5 | This work |
| Ag/FLG (PVDF) | 1M Na$_2$SO$_4$ | Stainless-steel | 729 (0.5) | 49.6 | 871 | This work |

* Two electrodes system

The electrochemical impedance spectroscopy (EIS) of Au/FLG and Ag/FLG electrodes was further applied to study the conductivity and charge transfer. The Nyquist plots of the studied electrodes measured over a frequency range of $10^{-2}$ to $10^5$ Hz are given in Fig. 13 (b). The plots are semicircle in the high-frequency region, a Warburg line in the intermediate frequency region, and almost a vertical line in the low-frequency region. The plot intercept and semicircle diameter relates to the equivalent series resistance (ESR) and charge transfer resistance ($R_{ct}$). The ESR value of Ag/FLG was found to be 3.1 Ω, slightly higher than that of Au/FLG (2.85 Ω). The charge transfer resistance ($R_{ct}$) attributed to the electrical conductivity of the electrode calculated for Ag/FLG from the semicircle diameter at the low-frequency region is 73.4 Ω, which is lower than the $R_{ct}$ of Au/FLG (128.6 Ω).

The capacitance retention of the composites was tested at a current of 5 mA, elucidating high capacitance stability at this current after 500 charges/discharge cycles, Fig. S6.

**Conclusions**

We have demonstrated the effective and smart use of industrial molasses co-product, first, as a surfactant with high hydrophilic-lipophilic balance (HLB) propriety in the synthesis of FLG, secondly as a stabilizer (and reductive agent ) for the metal over FLG surface. FLG is obtained with a very high yield by rapid and straightforward exfoliation method from graphite in water. Its presence in the synergy with molasses in the preparation of Au/Ag NPs leads to the diminished size of NP well dispersed over FLG flakes. Such prepared NPs/FLG composites showed efficient photocatalytic activity in the degradation of bisphenol (BisA) from wastewater in the presence of peroxymonosulfate (PMS) activator as well as relatively high supercapacity, the Ag/FLG being more performant than Au/FLG in both requests. The presented approach of using molasses waste co-product in the synthesis of FLG and its nanocomposites with metal based NPs creates new multiple pathways for further sustainable development of similar 2D nanomaterials and their hybrids. The synthesis method is incomparably advantageous to the production of rGO, the most common graphene-based material. The molasses - assisted preparation of nanocomposites with NPs can be extended to other metals, including non-noble metals.

**ASSOCIATED CONTENT**

**Supporting Information**

The following files are available free of charge (PDF) containing some of the molasses structures, TEM micrographs, and histograms of Au and Ag NPs size prepared in the absence of FLG, SEM micrographs of Ag/FLG, and Au/FLG photocatalysts after six tests, capacitance retention at 5 mA.

**AUTHOR INFORMATION**

**Corresponding Author**

Kamel Shoueir: kamel_rezk@nano.kfs.edu.eg, tel. +(33) 0749541481

Izabela Janowska: janowskai@unistra.fr, tel. +(33) 368852633

**Funding Sources**

This work was supported by the Egypt– France Cooperation program (STDF- IFE) n°42311 and CNRS/University of Strasbourg statutory funding.

**ACKNOWLEDGMENT**

Mahmoud Maher from the Institute of Engineering and Technology, Egypt, is acknowledged for his help with electrochemical measurements. Dr. Vasiliki Papaefthimiou (ICPEES) for performing XPS measurements. Dr. Walid Baaziz is acknowledged for carrying out TEM images at IPCMS (UMR 7504 CNRS-UDS), Strasbourg.